\documentclass[pra,showpacs,twocolumn]{revtex4}
\usepackage{amsmath,amsfonts,amssymb}
\usepackage{graphicx,float,calc}
\usepackage{color,bm}
\usepackage{ulem}
\usepackage[colorlinks,urlcolor=blue,citecolor=blue,linkcolor=blue]{hyperref}

\begin{document}
\author{Bin Wang$^{1}$}
\author{Zhen Zheng$^{1}$}
\author{Han Pu$^{2,3}$}
\thanks{hpu@rice.edu}
\author{Xubo Zou$^{1}$}
\thanks{xbz@ustc.edu.cn}
\author{Guangcan Guo$^{1}$}
\affiliation{$^1$Key Laboratory of Quantum Information, and Synergetic Innovation Center of Quantum Information \& Quantum Physics, University of Science and Technology of China, Hefei, Anhui, 230026, People's Republic of China}
\affiliation{$^2$Department of Physics and Astronomy, and Rice Center for Quantum Materials, Rice University, Houston, TX 77251-1892, USA}
\affiliation{$^3$Center for Cold Atom Physics, Chinese Academy of Sciences, Wuhan 430071, China}
\title{Effective $p$-wave interaction and topological superfluids in $s$-wave quantum gases}

\begin{abstract}
$P$-wave interaction in cold atoms may give rise to exotic topological superfluids.
However, the realization of $p$-wave interaction in cold atom system is experimentally challenging.
Here we propose a simple scheme to synthesize effective $p$-wave interaction in conventional $s$-wave interacting quantum gases. The key idea is to load atoms into spin-dependent optical lattice potential. Using two concrete examples involving spin-1/2 fermions, we show how the original system can be mapped into a model describing spinless fermions with nearest neighbor $p$-wave interaction, whose ground state can be a topological superfluid that supports Majorana fermions under proper conditions.
Our proposal has the advantage that it does not require spin-orbit coupling or loading atoms onto higher orbitals,
which is the key in earlier proposals to synthesize effective $p$-wave interaction in $s$-wave quantum gases,
and may provide a completely new route for realizing $p$-wave topological superfluids.
\end{abstract}
\pacs{03.75.Ss, 74.20.Fg, 03.65.Vf}
\maketitle

Topological superfluids have attracted tremendous attention in recent years. They serve 
as promising candidates to host Majorana fermions \cite{general-1,general-2,general-3}, which represents one of the most exotic quasiparticle states and may find applications in fault-tolerant quantum computing due to their non-Abelian nature.
It was first pointed out by Kitaev that a system of one dimensional (1D) spinless fermions with $p$-wave interaction
features a topologically nontrivial superfluid phase accompanied with zero-energy Majorana fermion states at boundaries
\cite{kitaev,read2000,jiang2011,diehl2011,lang2012,degottardi2013,cai2013,dumitrescu2013,wakatsuki2014}.
In ultra-cold atomic systems,
strong $p$-wave interactions have been realized directly via $p$-wave Feshbach resonances
in $^{40}$K and $^6$Li \cite{levinsen2008}.
Unfortunately, the $p$-wave Feshbach resonances in cold atoms are accompanied by large three-body inelastic collisional loss \cite{exp-1,exp-2}, and as a result, the lifetime of the Fermi gases is greatly reduced near the $p$-wave resonances \cite{exp-3,exp-4}.

On the other hand,
significant progress has been achieved in synthesizing effective $p$-wave interactions
in a conventional $s$-wave interacting Fermi gases.
The first progress along this line is based on synthetic spin-orbit coupling in atomic systems,
which is the generalization of the Kitaev model \cite{potter2010}.
In an $s$-wave Fermi gas, Rashba spin-orbit coupling can induce a spin-triplet pairing and thus create a chiral $p$-wave superfluid 
in a two dimensional (2D) system \cite{gorkov2001,zhang2008}.
However, experimental realization of spin-orbit coupling is accompanied by 
the large heating effect induced by the near-resonant Raman lasers
\cite{soc-1,soc-2,soc-3,soc-4,soc-5,soc-6}.
Another advance to engineer effective $p$-wave interaction
takes advantage of the odd parity due to $p$-orbital wave functions of atoms
\cite{p-wave-zoller,p-wave-buchler,p-wave-vincent}.
In these works, the Cooper pair with odd parity can arise by introducing pair hopping between two $s$-orbital atoms and one molecule,
or the $s$-wave interactions between one $s$- and one $p$-orbital atoms.
However this requires sophisticated optical lattice setup and demanding experimental manipulations to put atoms in $p$-orbitals \cite{p-orbital}.

In this Rapid Communication, different from earlier works,
we propose a new and very simple scheme for synthesizing $p$-wave superfluids in quantum gases
without involving either spin-orbit coupling or $p$-orbital atoms. The key ingredient in our proposal is a spin-dependent optical lattice potential consisting of two sublattices with spatial offset, and each sublattice traps a different atomic spin state. This configuration has already been realized in several cold atom laboratories. As we will show in detail below, this simple arrangement leads to an effective $p$-wave interaction whose strength is directly proportional to the intrinsic $s$-wave interaction strength, which can be tuned using $s$-wave Feshbach resonances. 
The simplicity of
our proposal makes it more feasible in experimental realization,
and may open up new avenues for studying topological superfluids
and the associated Majorana fermion states. In the following, we will illustrate our idea using two concrete examples involving spin-1/2 fermions in both a 1D and a 2D lattice geometry.

\begin{figure}[htbp]
\centering
\includegraphics[width=0.48\textwidth]{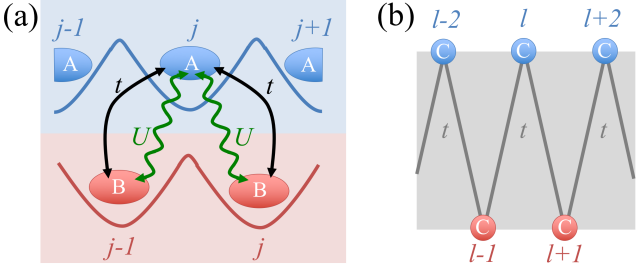}
\caption{(color online) (a) An illustration of the experimental proposal:
in the 1D optical lattice, fermions in hyperfine state $A$ (blue atom clouds) and $B$ (red atom clouds) experience the same lattice potential, but with an offset of half a lattice constant with respect to each other. As a result, each $A$ atom resides
in the center of two adjacent $B$ atoms, and vice versa.
The coupling between the two hyperfine states, as shown by the black solid lines,
is induced by an RF field.
The atom-atom interaction, as shown by the green wavy lines,
arises from the contact $s$-wave interaction between unlike spins.
(b) The 1D spinless chain model after mapping into the new index representation.}
\label{fig-model}
\end{figure}

\textit{1D lattice model.}
Our first example concerns a degenerate Fermi gas with two hyperfine states
(or \textit{pseudo-spins} denoted as $A$ and $B$, respectively) trapped in a 1D optical lattice potential.
Our proposed setup is illustrated in FIG.~\ref{fig-model}(a).
The lattice potential is spin-dependent and takes the form
\begin{equation}
V_A(x) = V_0\sin^2(k_L x) ~, \quad
V_B(x) = V_0\cos^2(k_L x) ~,
\label{eq-trap}
\end{equation}
where $k_L=\pi/a$ with the lattice constant $a$.
The recoil energy is defined as $E_R=h^2/2ma^2$ and 
will be chosen as the energy unit in the following.

Due to the spatial offset of sublattices $V_A$ and $V_B$, each $B$ site sits between two adjacent $A$ sites and vice versa,
as schematically illustrated in FIG.~\ref{fig-model}(a). We label the lattice sites such that the $j$-th site of $V_B$ is located between the $j$-th and the $(j+1)$-th sites of $V_A$.
In our proposal, fermions are deeply trapped in their respective sublattices and tunneling within each sublattice is negligible. However, the Wannier function of one sublattice site overlaps with the Wannier function of its nearest neighbors of the other sublattice. This establishes an interaction for the atoms trapped in the two sublattices. We can further induce a tunneling between the two sublattices by applying a radio-frequency (RF) field that drives a transition between the $A$ and $B$ states.
The Hamiltonian describing our lattice system takes the following form:
\begin{align}
H =& \sum\limits_{j} \Big[ -t (a_j^\dag b_{j-1} + a_j^\dag b_j + h.c. ) -
\mu( a_j^\dag a_j + b_j^\dag b_j ) \notag\\
&+ U ( a_j^\dag b_j^\dag b_j a_j
+ a_{j+1}^\dag b_j^\dag b_j a_{j+1}) \Big]~,
\label{eqn-Ho}
\end{align}
where $\mu$ is the chemical potential, 
$a_j$ and $b_j$ are annihilation operators of fermions on the $j$-th lattice site in states $A$ and $B$, respectively;
\begin{equation}
\begin{split}
t &= \Omega\int dx~W^*(x-{a}/{2})W(x) ~, \\
U &= g\int dx~|W(x-{a}/{2})\,W(x)|^2 ~,
\end{split}
\label{tu}
\end{equation}
characterize 
the RF field-induced \textit{inter}-sublattice tunneling amplitude and the interaction between nearest neighbors from different sublattices, respectively. Here $W(x)$ is the Wannier function of the lattice potential $V_A(x)$ or $V_B(x)$, $\Omega$ represents the RF field strength,
and $g$ is the contact $s$-wave interaction strength in free space.

\begin{figure}
\centering
\includegraphics[width=0.48\textwidth]{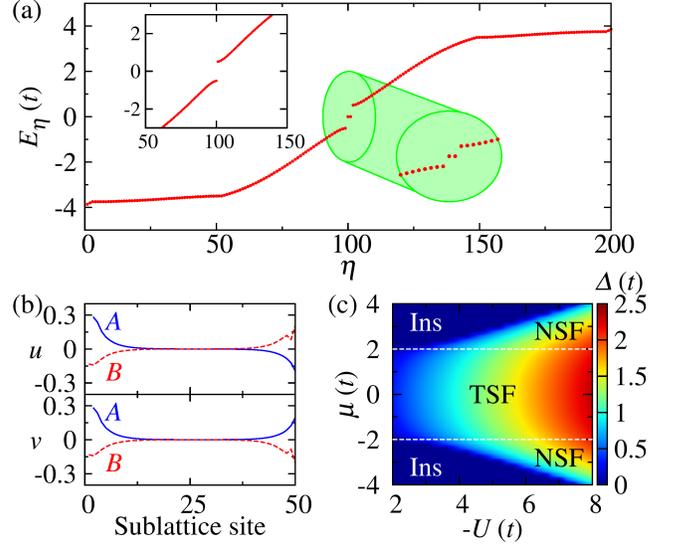}
\caption{(color online) (a) BdG quasi-particle excitation energies (in units of $t$)
for $V_0$=9.0$E_R$ at the number density $n=0.710$ ($\mu=1.5t$). Integer $\eta$ labels the quasi-particle states.
The zero-energy states are two-fold degenerate and gapped from the bulk.
The left inset shows the BdG quasi-particle excitation energies at $n=0.995$ ($\mu=2.5t$),
in which zero-energy states are absent.
Here we set $L=100$ and $U=-6.16t$.
(b) The spatial distribution of the zero-energy mode in the sublattice $A$ (blue solid lines) and $B$ (red dashed lines).
(c) Ground state phase diagram for Hamiltonian (\ref{eqn-Ho}): ``TSF/NSF" stands for the topological/nontopological superfluid state; and
``Ins" stands for the band insulator. The background color represents the magnitude of the order parameter.
$U$, $\mu$, and $\Delta$ are all shown in units of $t$.}
\label{fig-topo}
\end{figure}

To show that this system possesses nontrivial topological properties, we take the mean-field Bogoliubov-de Gennes (BdG) approach. Although the BdG theory is not expected to be quantitatively accurate for 1D fermionic systems, it can still capture many of the qualitative features, particularly the topological features of the system which we will focus on in this study. Here we assume that each sublattices contain $L/2$ sites, 
and introduce the superfluid order parameters $\Delta_{2j-1}=-U\langle b_ja_j\rangle$ and $\Delta_{2j}=-U\langle a_{j+1}b_j\rangle$.
The Hamiltonian (\ref{eqn-Ho}) can be diagonalized into the form $H = \sum_{\eta=1}^{L}\left(E_\eta\alpha^\dag_\eta\alpha_\eta-\frac{1}{2}\right) + \text{const.}$ by employing the BdG transformation
\begin{equation*}
a_j = \sum\limits_\eta \! \left( u_{2j-1}^\eta\alpha_\eta + v_{2j-1}^\eta\alpha_\eta^\dag \right) \,,\,
b_j = \sum\limits_\eta \! \left( u_{2j}^\eta\alpha_\eta + v_{2j}^\eta\alpha_\eta^\dag \right) \,,
\end{equation*}
with $\hat{u}_\eta=(u_1^\eta,\cdots,u_L^\eta)^T$ and $\hat{v}_\eta=(v_1^\eta,\cdots,v_L^\eta)^T$ satisfing the following BdG equations
\begin{equation}
\left(\begin{array}{cc}
\hat{h} & \hat{\Delta} \\
\hat{\Delta}^\dag & -\hat{h}
\end{array}\right)
\left(\begin{array}{c}
\hat{u}_\eta \\
\hat{v}_\eta^*
\end{array}\right)
=E_\eta
\left(\begin{array}{c}
\hat{u}_\eta \\
\hat{v}_\eta^*
\end{array}\right) ~,
\label{eq-real_BdG}
\end{equation}
where $E_n$ is the excitation energy for the $n$th quasiparticle state, $\hat{h}$ and $\hat{\Delta}$ are $L\times L$ matrices whose matrix elements are given by
\begin{equation*}
\begin{split}
\hat{h}_{ij} &= -\mu\delta_{ij}-t(\delta_{j,i-1}+\delta_{j,i+1})~,\\
\hat{\Delta}_{ij} &= -\Delta_i\delta_{j,i+1} + \Delta_{i-1}\delta_{j,i-1} ~,\\
\Delta_{j} &=- U\sum\limits_\eta \left[ u_j^\eta v_{j+1}^\eta\Theta(-E_\eta)+u_{j+1}^\eta v_j^\eta\Theta(E_\eta) \right]~,
\end{split}
\end{equation*}
where $\Theta$ is the Heaviside step function which describes the Fermi-Dirac distribution at zero temperature.

We apply an open boundary condition and solve the BdG Eqs.~(\ref{eq-real_BdG}) self-consistently. In FIG.~\ref{fig-topo}(a), we display quasiparticle spectrum $E_\eta$ for two sets of parameters. As one can see from the spectrum in the main figure, for $\mu<2t$, inside the superfluid gap between the positive-energy quasiparticle states and the negative-energy quasihole states, there exist two degenerate zero-energy modes that are separated from other modes by a finite gap. The wave functions of the zero-energy mode in the sublattices $A$ and $B$ are illustrated in FIG.~\ref{fig-topo}(b). As one can clearly see, they are well localized at the two ends of each sublattice chain. For larger atomic filling factor such that the chemistry potential $\mu$ exceeds $2t$, we find that the two mid-gap zero-energy states disappear, as shown in the inset of FIG.~\ref{fig-topo}(a). It characterizes the topological phase transition boundary as shown by white dashed lines in FIG.~\ref{fig-topo}(c). As it is well known that a 1D chain of spinless fermions with $p$-wave interaction hosts two zero-energy Majorana modes at the ends of the chain \cite{jiang2011}, we will pay particular attention to these two modes, and address the connection between our model and the spinless $p$-wave chain below.

Indeed our system can be mapped into the Kitaev model that describes a system of spinless fermions on a single 1D chain with nearest-neighbor $p$-wave interaction. This can be achieved by
redefining the fermion operator index with the mapping
\begin{equation}
a_j \rightarrow c_{2j}= c_{l}~, \quad b_j \rightarrow c_{2j+1}=c_{l+1} ~,
\end{equation} 
where $c$ represents the annihilation operator for a fictitious spinless fermion, in terms of which, the original Hamiltonian (\ref{eqn-Ho}) can be rewritten as 
\begin{equation}
H = -\sum\limits_{l} \Big[ t (c_l^\dag c_{l+1} +h.c.) +\mu c_l^\dag c_l 
- Uc_l^\dag c_{l+1}^\dag c_{l+1} c_l \Big] ~,
\label{BdG_n}
\end{equation}
which represents the Hamiltonian of a spinless fermion in a 1D lattice potential with the new lattice constant $\widetilde{a}=a/2$.
FIG.~\ref{fig-model}(b) illustrates this mapping, under which the hyperfine state $A$/$B$ of the original system corresponds to the even/odd site in the new representation. The $s$-wave interaction between the two sublattices is now mapped into an effective $p$-wave interaction between two identical fermions at two nearest-neighbor sites, with the same $U$ appearing in the original Hamiltonian (\ref{eqn-Ho}) characterizing the effective $p$-wave interaction strength.
Therefore, the existence of the zero-energy Majorana modes stems from the nontrivial topological properties of Hamiltonian (\ref{BdG_n}),
which is well known to generate the topological superfluid state when $\mu$ is below the critical value \cite{kitaev}.

\begin{figure}[htbp]
\centering
\includegraphics[width=0.48\textwidth]{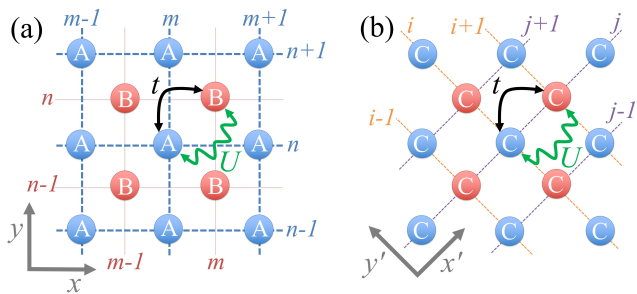}
\caption{(color online) (a) Illustration of the 2D lattice model.
An $A$ site (blue atom clouds) is surrounded by four $B$ sites(red atom clouds), and vice versa.
Similar to the 1D model,
the coupling between the two hyperfine states (black solid lines)
is induced by an RF field,
and the atom-atom interaction (green wavy lines)
arises from the contact $s$-wave interaction between unlike spins.
(b) The 2D spinless lattice model after mapping into the new index representation.}
\label{fig-model2D}
\end{figure}

\textit{2D lattice model.}
Our second example concerns a spin-1/2 Fermi gas in a 2D optical lattice.
The set-up is illustrated in FIG.~\ref{fig-model2D}(a).
We consider the 2D spin-dependent lattice potentials
\begin{equation}
\begin{split}
V_A(\bm{r}) &= V_0 \sin^2(k_Lx) + V_0 \sin^2(k_Ly) ~, \\
V_B(\bm{r}) &= V_0 \cos^2(k_Lx) + V_0 \cos^2(k_Ly) ~.
\end{split}
\label{eq-trap2D}
\end{equation}
In such a 2D optical lattice,
each fermions in the pseudo-spin $A$ state resides in the middle of four $B$ fermions, and vice versa.
The \textit{inter}-sublattice tunneling is again induced by an RF field,
while the \textit{intra}-sublattice tunneling is made to be negligible.
As in our previous example, each $A$ atom can interact with four nearest-neighbor $B$ atoms
by virtue of the spatial overlap of the Wannier functions of the two sublattice trap potentials $V_A(\bm{r})$ and $V_B(\bm{r})$.
The Hamiltonian in the tight-binding approximation is therefore expressed as
\begin{align}
H =& \sum\limits_{\langle mn, m'n' \rangle} \big(
-ta_{mn}^\dag b_{m'n'} + Ua_{mn}^\dag b_{m'n'}^\dag b_{m'n'} a_{mn} \big) \notag\\
&-\mu\sum\limits_{mn}( a_{mn}^\dag a_{mn} + b_{mn}^\dag b_{mn} ) ~.
\label{eqn-Ho2D}
\end{align}
Here $\sum_{\langle mn, m'n' \rangle}$ accounts for the summation between $(m,n)$-th site and its four nearest-neighbor sites in the other sublattice.
Similar to the 1D model, the \textit{inter}-sublattice tunneling $t$
and interaction amplitude $U$ are given in Eqs.~(\ref{tu}), only that here the integrals are taken on the 2D space. 

\begin{figure}[htbp]
\centering
\includegraphics[width=0.48\textwidth]{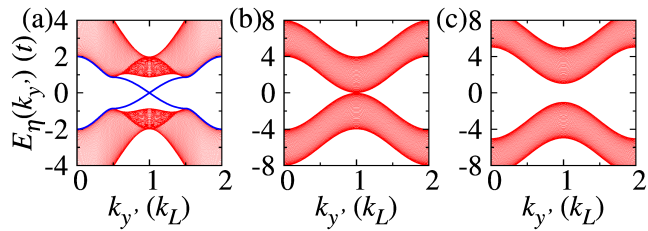}
\caption{(color online) Topological edge states in a 2D strip.
Tuning the chemistry potential $\mu$, the 2D model exhibits a phase transition from (a) the chiral $p$-wave superfluid state
to (c) a band insulator.
We fix $U=-6.31t$, and set $\mu=2.0t$ ($n=0.790$) for (a), $\mu=4.0t$ ($n=$1.000) for (b), and $\mu=5.0t$ ($n=$1.000) for (c).
$E_\eta(k_{y'})$ is shown in units of $t$ while $k_{y'}$ in units of $k_L$.}
\label{fig-Ek2D}
\end{figure}

Following the similar mean-field BdG approach, we introduce the superfluid order parameters
$\Delta_{x',2i-1}=-U\langle b_{mn}a_{mn}\rangle$, $\Delta_{x',2i}=-U\langle a_{m+1,n+1}b_{mn}\rangle$,
$\Delta_{y',2j-1}=-U\langle b_{m-1,n}a_{mn}\rangle$, and $\Delta_{y',2j}=-U\langle a_{m-1,n+1}b_{m-1,n}\rangle$.
With an open (periodic) boundary condition in the $x$ ($y$) direction,
the self-consistently solved BdG equations yield the quasiparticle spectrum shown in FIG.~\ref{fig-Ek2D}.
We find that the superfluid system supports gapless chiral edge modes when the chemical potential $\mu$ is below the critical value $4t$.
Tuning $\mu$ across $4t$, the bulk band gap closes and reopens,
revealing a topological phase transition from the topologically nontrivial superfluid state ($\mu <4t$) to a topologically trivial band insulator ($\mu >4t$).
Meanwhile for the topological superfluid, there exists a relative phase factor $e^{\pm i\pi/2}$ between $\Delta_{x'}$s and $\Delta_{y'}$s.
To understand more clearly the topological properties of the system,
we rotate the $x$-$y$ plane by 45$^{\rm o}$ anticlockwise,
and redefine the fermion operator index by mapping the original spin-1/2 system into a spinless Fermi gas with nearest-neighbor $p$-wave interaction, as shown in FIG.~\ref{fig-model2D}(b).
With a periodic boundary condition along both $x$ and $y$ directions,
the momentum space BdG Hamiltonian of the mapped system can be written as
\begin{equation}
H_\mathrm{BdG}(\bm{k})= 
\left(
\begin{array}{cc}
\epsilon(\bm{k}) & \Delta_{\bm{k}} \\
\Delta_{\bm{k}}^* & -\epsilon(\bm{k})
\end{array}
\right) ~,
\label{BdG_k2D}
\end{equation}
where $\epsilon(k)=-2t\cos(k_{x'}\widetilde{a})-2t\cos(k_{y'}\widetilde{a})-\mu$ is the single-particle dispersion
with a new lattice constant $\widetilde{a}=a/\sqrt{2}$,
and $\Delta_{\bm{k}}=i\Delta_{x'}\sin(k_{x'}\widetilde{a}) +i\Delta_{y'}\sin(k_{y'}\widetilde{a})$.
Our calculation based on the mapped Hamiltonian (\ref{BdG_k2D}) shows that the ground state is a superfluid state with $\Delta_{x'}=\pm i\Delta_{y'}$,
which confirms the obtained result based on the original model in the real space.
It indicates that for $\mu <4t$, the ground state of this 2D system is the spontaneously time-reversal broken
chiral $p_x\pm ip_y$ superfluid state.
That the $p_x\pm ip_y$ fermionic superfluid supports zero-energy Majorana modes
has been well studied in earlier works \cite{p-wave-buchler,p-wave-vincent}.

\textit{Experimental feasibility.}
Our proposal can be readily realized using current technology.
The key ingredient is the spin-dependent lattice potential in Eqs.~(\ref{eq-trap}) and (\ref{eq-trap2D}).  
Here we put forward two ideas for its implementation.
Perhaps the most straightforward way is to use a standing wave laser field whose frequency is properly tuned such that it is blue detuned for one spin state and red detuned for the other. In this way, the lattice potentials for the two spin states will be 180$^{\rm o}$ out of phase, exactly realizing the potentials (\ref{eq-trap}) and (\ref{eq-trap2D}). The second idea is to exploit the polarization of the laser fields. For example, in the lin $\perp$ lin configuration, where two counter-propagating linearly polarized traveling waves with polarization perpendicular to each other are employed, two standing waves with $\sigma^\pm$ polarization and alternating maxima/minima will be formed. This idea has been implemented in several experiments to create spin-dependent lattice potentials \cite{spin-dependent-1,spin-dependent-2}.
As a concrete example, let us consider $^{40}$K and
choose the hyperfine state $|F,m_F\rangle=|9/2,-9/2\rangle$ as state $A$ and
$|9/2,-7/2\rangle$ as $B$.
Based on the second idea, we can realize a lattice with depth $V_0=9E_R$ using lasers with wavelength $\lambda=1064\,$nm.
The recoil energy of such an optical lattice is $E_R=h^2/2m\lambda^2 \approx 210\,$nK. By choosing a proper RF field strength, we can tune $\Omega=1.0E_R$. Using the standard technique of Feshbach resonance, we can set $g=-5.0E_R$. These choices lead to
the system parameters used in FIGs.~\ref{fig-topo} and \ref{fig-Ek2D}
with
the tunneling energy $t\approx0.21E_R\approx 44\,$nK,
and the interaction strength is $U\approx -1.3E_R\approx -6t$.

In summary, we have proposed a method to realize $p$-wave superfluids
and its supported Majorana fermion states in $s$-wave spin-1/2 Fermi gases with properly engineered spin-dependent lattice potential. Our proposal possesses the following important advantages: (1) It employs only standard techniques that have been well demonstrated and are readily implementable in practice. In particular, it requires no spin-orbit coupling and no need to load atoms onto high orbitals. (2) Without the need of the $p$-wave Feshbach resonance, the effective $p$-wave interaction strength in our proposal can be made to be very strong as it is directly proportional to and on the same order of the intrinsic $s$-wave interaction strength. 

Finally we comment that even though we used spin-1/2 fermions in our examples, the same idea can be applied to spin-1/2 bosons, in which case the system can be mapped into a system of spinless bosons with nearest neighbor interaction that may host novel quantum phases, in particular, the supersolid phase that simultaneously possesses crystalline and superfluid orders
\cite{bose1,bose2,bose3,bose4,bose5}. Furthermore, it can be generalized to higher spatial dimensions where more exotic phases and effective higher partial-wave interactions can be expected.

\textit{Acknowledgments.}
B.W., Z.Z., X.Z. and G.G. are supported by National Natural
Science Foundation of China (Grants No. 11074244 and No. 11274295),
and National 973 Fundamental Research Program (No. 2011cba00200).
H.P. acknowledges support from the US NSF and the Welch Foundation (Grant No. C-1669).

\end{document}